\documentclass[aps,prd,showkeys,nofootinbib,superscriptaddress,12pt]{revtex4-2}

\usepackage{color}
\definecolor{darkgreen}{rgb}{0,0.35,0}
\newcommand{\rc}{\textcolor{red}}

\usepackage{hyperref}
\usepackage{amsfonts}
\usepackage{amsmath}
\usepackage[noabbrev]{cleveref}
\usepackage{subfig}
\usepackage{mathtools,slashed,tensor}
\usepackage[font=scriptsize,labelfont=bf]{caption}
\usepackage{float}
\usepackage[top=2cm, bottom=2cm, left=3cm, right=3cm]{geometry}
\usepackage{tikz-cd}
\usepackage{siunitx}

\begin{document}
\newcommand{\eg}{{\it e.g.}}
\newcommand{\etal}{{\it et. al.}}
\newcommand{\ie}{{\it i.e.}}
\newcommand{\be}{\begin{equation}}
\newcommand{\dd}{\displaystyle}
\newcommand{\ee}{\end{equation}}
\newcommand{\bea}{\begin{eqnarray}}
\newcommand{\eea}{\end{eqnarray}}
\newcommand{\bef}{\begin{figure}}
\newcommand{\eef}{\end{figure}}
\newcommand{\bce}{\begin{center}}
\newcommand{\ece}{\end{center}}
\newcommand \bed {\begin{displaymath}}
\newcommand \eed {\end{displaymath}}
\newcommand \lan {\langle}
\newcommand \ran {\rangle}
\newcommand{\bi}{\bibitem}
\newcommand{\eit}{\end{itemize}}
\newcommand{\Z}{\mathbf{Z}}
\newcommand{\R}{\mathbf{R}}
\newcommand{\1}{\mathbf{1}}
\newcommand{\eps}{\epsilon}
\newcommand{\beps}{\bar\epsilon}
\newcommand{\bpsi}{\bar\psi}
\newcommand{\blambda}{\bar\lambda}
\newcommand{\bsigma}{\bar\sigma}
\newcommand{\dalpha}{\dot\alpha}
\newcommand{\dbeta}{\dot\beta}
\newcommand{\dgamma}{\dot\gamma}
\newcommand{\ddelta}{\dot\delta}
\newcommand{\lra}{\leftrightarrow}
\newcommand{\la}{\leftarrow}
\newcommand{\ra}{\rightarrow}
\newcommand \dsl {\not\!\partial}
\newcommand \dslbar {\not\!\bar\partial}
\newcommand{\bgamma}{\mbox{\boldmath{$\gamma$}}}
\newcommand{\ba}{\mbox{\boldmath{$a$}}}
\newcommand{\bA}{\mbox{\boldmath{$A$}}}
\newcommand{\bD}{\mbox{\boldmath{$D$}}}
\newcommand{\bw}{\mbox{\boldmath{$w$}}}
\newcommand{\bomega}{\mbox{\boldmath{$\omega$}}}
\newcommand{\bphi}{\mbox{\boldmath{$\phi$}}}
\newcommand{\bOmega}{\mbox{\boldmath{$\Omega$}}}
\newcommand{\bGamma}{\mbox{\boldmath{$\Gamma$}}}
\newcommand{\bO}{\mbox{\boldmath{$O$}}}
\newcommand{\bW}{\mbox{\boldmath{$W$}}}
\newcommand{\bu}{\mbox{\boldmath{$u$}}}
\newcommand{\bU}{\mbox{\boldmath{$U$}}}
\newcommand{\bV}{\mbox{\boldmath{$V$}}}
\newcommand{\bS}{\mbox{\boldmath{$S$}}}
\newcommand{\bN}{\mbox{\boldmath{$N$}}}
\newcommand{\bB}{\mbox{\boldmath{$B$}}}
\newcommand{\vt} [1] {\textit{\textbf{#1}}}
\newcommand{\deriv}[1]{\frac{\partial}{\partial #1}}
\newcommand{\bgar}{\begin{eqnarray}}
\newcommand{\enar}{\end{eqnarray}}
\newcommand{\eq}[1]{(\ref{eq:#1})}
\newcommand{\ket}[1]{ \left. | #1 \right\rangle }
\newcommand{\bra}[1]{\left\langle  #1 | \right.}
\newcommand{\omp}{\omega_{p}}
\newcommand{\omc}{\omega_{c}}
\newcommand{\dtp}{\delta_{p}}
\newcommand{\dtc}{\delta_{c}}
\newcommand{\mean}[1]{\left\langle #1 \right\rangle}
%
\newcommand{\beqar}[1]{\begin{eqnarray}\label{#1}}
\newcommand{\eeqar}{\end{eqnarray}}
\newcommand{\bas}{\bar{\alpha}_s}
\newcommand{\un}{\underline}
\newcommand{\lag}{\mathcal{L}}
\newcommand{\vac}{|\epsilon_\mathrm{v}|}
\newcommand{\jpsi}{{J/\Psi}}
\newcommand{\lqcd}{\Lambda_\mathrm{QCD}}
\newcommand{\stackeven}[2]{{{}_{\displaystyle{#1}}\atop\displaystyle{#2}}}
\newcommand{\im}{\mathrm{Im}}
\newcommand{\arcsinh}{\mathrm{arcsinh}}
\newcommand{\arctanh}{\mathrm{arctanh}}

\title{Hunting Quantum Gravity with Analogs: \\
the case of High Energy Particle Physics}

\author{Paolo Castorina}
\affiliation{Istituto Nazionale di Fisica Nucleare, Sezione di Catania, I-95123 Catania, Italy}
\affiliation{Faculty of Mathematics and Physics, Charles University, V Hole\v{s}ovi\v{c}k\'ach 2, CZ-18000 Prague 8, Czech Republic}

\author{Alfredo Iorio}
\affiliation{Faculty of Mathematics and Physics, Charles University, V Hole\v{s}ovi\v{c}k\'ach 2, CZ-18000 Prague 8, Czech Republic}

\author{Helmut Satz}
\affiliation{Fakult\"{a}t f\"{u}r Physik, Universit\"{a}t Bielefeld, D-33501 Bielefeld, Germany}

\email{castorina@ct.infn.it}
\email{iorio@mff.cuni.cz}
\email{satz@physik.uni-bielefeld.de}


\date{\today}

\begin{abstract}
In this review we collect, for the first time in one paper, old and new results and future perspectives of the research line that uses hadron production, in high-energy scattering processes, to experimentally probe fundamental questions of quantum gravity. The key observations, that ignited the link between the two arenas, are the so-called ``color-event horizon'' of quantum chromodynamics, and the enormous (de)accelerations involved in such scattering processes: both phenomena point to the Unruh (and related Hawking) type of effects. After the first pioneering investigations of this, such research went on and on, including studies of the horizon entropy and other ``black-hole thermodynamical'' behaviors, which incidentally are also the frontier of the analog gravity research itself. It is stressed in various places here that the \textit{trait d'union} between the two phenomenologies is that in both scenarios, hadron physics and black hole physics, ``thermal'' behaviors are more easily understood not as due to real thermalization processes (sometimes just impossible, given the small number of particles involved), but rather to a stochastic/quantum entanglement nature of such temperature.
Finally, other aspects, such as the self-critical organizations of hadronic matter and of black-holes, have been recently investigated. The results of those investigations are also summarized and commented upon here. As a general remark, this research line shows that indeed we can probe quantum gravity theoretical constructions with analog systems that are not confined to belong only to the condensed matter arena. This is as it must be.
\end{abstract}

\maketitle


\section{Introduction}

Analogs have reached a level of maturity, both on the theoretical modelling side, see, e.g., \cite{Barcelo2005}, and on the experimental side, see, e.g., \cite{MunozdeNova2019}, that might bring them at the forefront of the experimental search for signatures of quantum gravity (QG) or, in general, of the theoretical research in fundamental high energy physics, see, e.g., the contribution \cite{UniverseGraphene} to this Issue.

Two are the obstacles on this way: first, the skepticism of a large part of the theoretical community, that still does not trust analogs as a way to test the fundamental ideas; second, the need of a new era in the
analog enterprise, namely to reach \textit{dynamical} effects, rather than \textit{kinematical} effects.

Here we describe a line of research, initiated in \cite{KHARZEEV2005316,Castorina2007}, that addresses both problems. It is focussed on a specific high energy scenario, where the effects of a large acceleration are evident, and much of the subsequent work has been carried on to understand the meaning of entropy in this context and its relation to black-hole (BH) entropy, that is a typical dynamical issue (recall, e.g., that Wald's formula relates entropy to the action \cite{Wald1993}).

The reproduction of aspects of gravitational physics, both classical and quantum, by means of analogs is mainly based on condensed matter systems. Examples range from lasers \cite{PisinChen1999,Habs2006,Habs2008,Kim2017,LOR2001} (see also the contribution \cite{UniverseLaser} to this Issue) to water-waves \cite{Leonhardt2019}, from Bose-Einstein condensates \cite{MunozdeNova2019} to graphene \cite{i2,weylgraphene,iorio2012,iorio2014,iorio2015,ipfirst,ip3,ip5,egypt2018,ip4,GUPBTZ} and more \cite{Barcelo2005}.

In particular the detection of some form of the Unruh phenomenon \cite{haw0,bill,crispino} has been proposed in various set-ups \cite{PisinChen1999,Habs2006,Habs2008,Kim2017,iorio2012,iorio2014,Leonhardt2019,opt1,opt2,opt3}.
However, in many of the proposed analog systems, the Unruh temperature
\be\label{UnruhT}
T_U  = \frac{\hbar a}{2\pi c k_B}
\ee
is still too small \cite{KHARZEEV2005316} for a direct experimental verification, as one sees that $1 {\rm m}/{\rm s}^2 \to \sim 4 \times 10^{-21}$K. In (\ref{UnruhT}) $a$ is the uniform acceleration and we explicitly kept the Planck constant, the speed of light and the Boltzmann constant to ease the units conversion. In the following we shall set to one $\hbar$,$c$ and $k_B$.

Some encouraging results come from femtosecond laser pulses that can produce an acceleration $a \simeq 10^{23} {\rm m}/{\rm s}^2 $ \cite{Kim2017}, with associated Unruh temperature $T_U \sim 400$K. On the other hand, the enormous accelerations (or decelerations) produced in relativistic heavy ion collisions, $a \simeq 4.6 \times 10^{32} {\rm m}/{\rm s}^2$, have associated Unruh temperatures many orders of magnitude bigger, $T_U \sim 1.85 \times 10^{12}$K. A simple units conversion shows that this is nothing else than the hadronization temperature $T_h$
\be
T_U  \sim 160 {\rm MeV} \sim T_h \,.
\ee
This fact triggered the investigation of hadron production, in high energy collisions, as a manifestation of the Unruh phenomenon in Quantum Chromodynamics (QCD) \cite{KHARZEEV2005316,Castorina2007}.

Of the latter we shall discuss in this paper, by reviewing why such interpretation is very natural, commenting on the various ramifications and speculating on the possible future directions. In other words, we shall here elaborate on which aspects of this QCD phenomenology can be taken as viable analogs of specific aspects of QG.

The underlying idea behind the latter analogies is based on quark confinement as a phenomenon where an ``horizon'' (sometimes called ``color horizon'', see, e.g., \cite{recami,salam}) hides those degrees of freedom to any observer, and only quantum (tunneling) effects could explain a radiation phenomenon \cite{KHARZEEV2005316}. This is a non-perturbative quantum phenomenon, related to the chromo-magnetic properties of the QCD vacuum (see for example ref.\cite{digiacomo}), producing the squeezing of the chromoelectric field in quark-antiquark strings, with a constant energy density. Let us comment a bit more on this.

Quark confinement can be described by a potential that grows linearly at large distances, $V = \sigma r$. This corresponds to a constant acceleration, henceforth the Rindler spacetime is the appropriate framework for this phenomenon. As well known, the Rindler metric is equivalent to the near horizon approximation of some BH metric, with the acceleration equal to the surface gravity, $k$. Therefore, the \textit{local} correspondence between a linear potential and the near-horizon dynamics of a BH is a very strong analogy.

This is another perspective on why quark confinement can be related to a ``color horizon'' \cite{recami,salam}, that is both something hiding the color degrees of freedom and a Rindler horizon, in turn associated to some specific BH (in \cite{grumi} are some proposals of specific BHs that could account for this specific scenario). On the other hand, Hawking radiation is a quantum phenomenon associated with tunneling and pair-creation, near the event horizon \cite{parwilc,vanzo}. This is a clear dynamical correspondence with the string breaking and quark-antiquark pair-creation, in the final process of the mechanism leading from color degrees of freedom till the formation of hadrons.

Finally, another delicate dynamical issue. The entropy associated to a ``color event horizon'' is necessarily an entanglement entropy between quantum field modes on the two sides of the horizon. As well known, such an entropy follows an area law \cite{sre,tera,alfgae}, just like the entropy of a BH \cite{bek,haw1}, when logarithmic corrections are not included, or the entropy of a Rindler horizon \cite{laflamme}. Even though it is still an open question whether entanglement entropy alone could account for the whole of BH entropy, this is yet another argument that strengthen the analogy between the two systems. Furthermore, in such QCD environments the entropy is a quantity routinely considered, e.g., in (quantum) statistical models. Henceforth, we have measurable and natural candidates for quantities that can play the role of a BH entropy. As mentioned earlier, this is a very important milestone to move analogs to the next era, that is, the possibility to reproduce BH thermodynamics, with its intriguing fundamental open questions, such as the information paradox. For sake of completeness let us recall that the general Page approach to the calculation of the entanglement entropy of an evaporating BH \cite{Page1993a} has been successfully applied to gluon shadowing in deep inelastic scattering \cite{shadowing}, following the proposal of \cite{dima1}.

In this review paper we shall collect in one place, for the first time, the most important old and new results of this line of research, and shall comment and discuss them. The paper is organized as follows. In Section \ref{ReviewUnruhHawking} we recall the main features of the Unruh effect, and of the related BH physics, using those descriptions of the effect that make easier the link with hadron physics that we want to disclose. That Section is also important for setting the notation. In Section \ref{HadronicPhenomenology} we then recollect three well known aspects of the phenomenology of hadrons, that will be scrutinized, using the analogies and links with gravitational physics, in the Sections that follow. The correspondence is such that: the hadronic phenomena described in Subsection \ref{StatHadrModel} are re-interpreted as gravity analogs in Section \ref{SecStatHadrModel}; the hadronic phenomena described in Subsection \ref{TransverseMomentum} are re-interpreted as gravity analogs in Section \ref{SecTransverseMomentum}; finally, the hadronic phenomena described in Subsection \ref{SelfOrganization} are re-interpreted as gravity analogs in Section \ref{SecSelfOrganization}. We close the review with our conclusions in Section \ref{Conclusions}.

\section{Accelerated observers and near BH horizon observers}\label{ReviewUnruhHawking}

In this Section we recall the main features of the Unruh effect, and of the related BH physics, that will mostly be used in the realizations in hadronic physics, that we discuss later. In particular, we first discuss the interplay between pair-production, tunneling and the Unruh effect. We then mention the correspondence between near horizon BH metric and Rindler metric, and the area law obeyed by BH entropy.

Let us start by discussing the Unruh effect and its relation to tunneling and pair-production. For this part we shall follow \cite{KHARZEEV2005316}.

Consider the action, $A$, of a particle of mass $m$, subject to a constant force derived from a potential $\varphi(x)$:
\begin{equation} \label{actfield}
A \, = - \, \int \, \, ( m \, ds \, +\,\, \varphi \,dt ) \,.
\end{equation}
For a constant force, the one-dimensional (1D) potential is $\varphi=- \sigma x$ modulo an additive constant and the  equations of motion of the particle are
\be \label{eqmotEM}
\frac{d p_x}{dt} \, = \sigma ,\quad \frac{d p_\bot}{dt}\,=\,0\,.
\ee
Using $ds^2=(1-v^2(t))\,dt^2$ and the equations of motion, one can evaluate the action $A$ \cite{KHARZEEV2005316}
\bea
A(\tau) &=& \int^\tau\,dt\,(-\,m\,\sqrt{1\,-\,v(t)^2}\,+\,\sigma x(t)) \nonumber \\
&=& -\,\frac{m}{a}\, {\rm arcsinh} (a \tau) \, + \,
\frac{\sigma}{2\,a^2}\,[a\,\tau\,(\sqrt{1\,+\,a^2\,\tau^2}\,- 2)\,+\,\, {\rm arcsinh} (a\,\tau)] + \mathrm{const}\,.
\eea

In the quantum theory, the particle has a finite probability to be found under the
potential barrier, $\sigma x$, in the classically forbidden region. Mathematically, this
comes about because the action $A$, being an analytic function of $\tau$, has an imaginary part
\be
\label{impart}
A(\tau)\,=\,\frac{m\,\pi}{a}\,-\,\frac{\sigma \pi}{2\,a^2}\,=\,\frac{\pi\,m^2}{2 \sigma},.
\ee
which corresponds to the motion of a particle in Euclidean time, $t_E$, with Euclidean trajectory
\be
\label{euctraj}
x(t_E)\,=\,a^{-1}\, \left(\,\sqrt{1\,-\,a^2\,t_E^2}\,-\,1\,\right) \,,
\ee
bouncing between the two identical points $x_a=-a^{-1}$ at $t_{E,a}=-a^{-1}$ and
$x_b=-a^{-1}$ at $t_{E,b}=a^{-1}$, and the turning point $x_a = 0$ at $t_{E,a}=0$.

In the quasi-classical approximation, the rate of tunneling under the potential barrier is thus given by
\be
\label{gamma1}
\Gamma_{vac\rightarrow m}\,\sim\,e^{-2\, {\rm Im} A}\,=\,e^{-\frac{\pi\,  m^2}{\sigma}}\,,
\ee
which gives the probability to produce a particle and its antiparticle (each of mass $m$) out of the vacuum, under the effects of a constant force $\sigma$. The ratio of the probabilities to produce states of masses $M$ and $m$ is then
\be\label{ratprob}
\frac{\Gamma_{vac\rightarrow M}}{\Gamma_{vac\rightarrow m}}\,=\,e^{-\frac{\pi\,  (M^2\,-\,m^2)}{\sigma}}\,.
\ee

The relation (\ref{ratprob}) had a double interpretation, in terms of both the Unruh and the Schwinger effects, see e.g. \cite{Parentani,Gabriel,Narozhny} and references therein. Indeed, consider a detector with quantum levels $m$ and $M$, moving with a constant acceleration. Each level is accelerated differently, however if the splitting is not large, $M-m \ll m$, we can introduce the average acceleration of the detector
\be\label{averacc}
\bar a\,=\,\frac{2\sigma}{M\,+\,m}\,.
\ee
Substituting (\ref{averacc}) into (\ref{ratprob}), we arrive at
\be\label{ratun}
\frac{\Gamma_{vac\rightarrow M}}{\Gamma_{vac\rightarrow m}}\,=\,e^{\frac{2\,\pi\,(M\,-\,m)}{\bar a}}\,.
\ee
This expression is reminiscent of the Boltzmann probabilistic weight in a heat bath, with an effective temperature, $T = \bar a/2 \pi$. This is the Unruh effect.

A similar study of the Unruh radiation, as tunneling through a barrier by WKB–like methods has been carried out in \cite{singleton}. A more rigorous derivation of the Unruh effect
can be given by recalling that the uniformly accelerated detector in Minkowski space is equivalent to the inertial detector in the Rindler space. The vacuum in Minkowski space is related to the vacuum in the Rindler space by a nontrivial
Bogoliubov transformation, which shows that the Rindler vacuum is populated with a thermal radiation of temperature $T=a/2\pi$ (for a review see \cite{crispino}).

Let us now focus on another aspect of the Hawking-Unruh phenomenon that is crucial for the analogy between quark confinement and the physics of curved spacetime, that we shall discuss later. That is the correspondence between Rindler metric and the near horizon approximation of a BH metric.

The Schwarzschild metric for a BH of mass $M$, in radial coordinates, is given by
\be
ds^2 = f(r)~\!dt^2 - f(r)^{-1}~\!dr^2 - r^2[d\theta^2 + \sin^2\theta d\phi^2],
\label{metric}
\ee
with
\be
f(r) =\left(1 - {2GM\over r}\right) \,.
\ee
The equation $f(r) = 0$ sets the Schwarzschild radius, $R_S$, as the radius of the spherical event horizon
\be
R_S= 2~\!G~\!M.
\ee
This means that $M(R_S) = (2G)^{-1}\ R_S$, that is a linear law for the BH mass. This is particularly interesting if one notices that an analogous behaviour is enjoyed by the confining potential of the strong interactions.

Making in Equation (11) the coordinate transformation \cite{tera}
\be
\eta = {\sqrt{f(r)} \over \kappa},
\ee
where the surface gravity $\kappa$ is given by
\be
\kappa = {1\over 2}~\!\left({\partial f \over \partial r}\right)_{r=R_S},
\ee
one obtains, for $r \to R_S$, the BH metric in the near horizon approximation
\be
ds^2 =
 \eta^2 \kappa^2 dt^2 - d\eta^2 - R^2
(d\theta^2 + sin^2\theta d\phi^2) \,.
\label{R-bh-metric}
\ee

To compare the previous result with the Rindler metric of a constantly accelerated observer, let us recall the relation among Rindler coordinates, ($\xi,\tau$) and the Minkowski coordinates ($x,t$)
\be
x = \xi \cosh a\tau \,, ~~~~~ t= \xi \sinh a \tau,
\label{rindler}
\ee
where $a=\sigma/m$ denotes the acceleration in the instantaneous rest frame of $m$, and $\tau$ is the proper time. With these, the metric of such an accelerating system, in spherical coordinates is
\be
ds^2 = \xi^2 a^2 d\tau^2 - d\xi^2 - \xi^2 \cosh^2 a \tau (d\theta^2
 + \sin^2 \theta d\phi^2) \,.
\label{R-metric}
\ee
Comparing Equations (\ref{R-metric}) and (\ref{R-bh-metric}), it is evident that the system in uniform acceleration is the same as a system near a spherical BH horizon, provided we identify the surface gravity $\kappa$ with the acceleration $a$.

The final topic of this Section will be the entropy and its area law, a feature common to BH and constantly accelerated systems.

The gravitational entropy is related to the existence of a horizon which forbids an observer to acquire knowledge or what is happening beyond it. In a way, it could be seen as a measure of the ignorance on the fate of matter (and space) degrees of freedom that contributed making the BH.

As well known, such entropy obeys the Hawking-Bekenstein area law \cite{bek,haw1}:
\be\label{SBekHaw}
S_{\rm BH} = \frac{1}{4} \frac{A}{\ell_P^2} \,,
\ee
where $\ell_P= \sqrt{G}$ is the Planck length, and, for a  Schwarzschild BH,  $A = 4 \pi R_S^2$. Once more the only parameter of interest is the mass of the BH, $M \sim R_S$.

On the other hand, it is also well known that the access to the degrees of freedom describing an accelerated observer are also restricted by a horizon, the Rindler horizon. Therefore, the entropy of the, so called, Rindler wedge has been evaluated along similar lines as for the BH. The computation was performed long time ago \cite{laflamme}, and it turns that $S = (1/4) \, (A_a/\ell_P^2)$, where $A_a$ is the area of a surface of constant Rindler spatial coordinate, $x$, and proper time, $\tau$.  If $y, z$ are the Minkowski coordinates (we suppose that the acceleration is along the $x$ axis), the entropy is actually infinite, however, an entropy \textit{density}, per unit area, can be defined for this spacetime.

Finally, we recall here another well known result, namely that the entanglement (hence quantum) entropy of a bipartite system (which include both the Rindler and thee BH cases just discussed, due to their event horizons) also obeys an area law. This has been shown in various quantum field theoretical set-ups, see, e.g., \cite{sre,tera,alfgae}. In the following we shall discuss also this feature, when we shall focus on the phenomenological implications of the area law for the entropy in hadron production.


\section{Aspects of the hadron production in high energy collisions that have a gravity analog}\label{HadronicPhenomenology}

Now we schematically recall those results of the phenomenological analysis of high energy collisions data, that have been interpreted in analog gravity scenarios. As announced in the Introduction, each of the following three Subsections will then be reconsidered, in the light of the gravity analog, in a separate dedicated Section.

\subsection{Statistical Hadronization Model}\label{StatHadrModel}

There is an abundant multihadron production in high-energy collisions, starting from the electron-positron annihilation, and then in the proton-proton, proton-nucleus and nucleus-nucleus scattering processes.
The relative rates of the secondaries thus produced are well accounted for by an ideal gas of all hadrons and all hadronic
resonances, at fixed temperature $T$ and baryochemical potential $\mu_B$. This is known as the Statistical Hadronization Model (SHM)  \cite{beca1,cley,PBM2}. There is one, well known, caveat though. The strangeness production one finds is
reduced with respect to the rates predicted by the SHM . This suppression can, however, be taken into account by one further parameter, $0 < \gamma_s \leq 1$, if the predicted rate for a hadron species containing $\nu=1,2,3$ strange quarks is suppressed by
the factor $\gamma_s^{\nu}$ \cite{LRT}.

To describe such a resonance gas, the basic tool one needs is the grand-canonical partition function for an ideal gas at temperature $T$ in a spatial volume $V$
\be
\ln Z(T) = V \sum_i {d_i \gamma_s^{\nu_i}\over (2\pi)^3}~\! \phi(m_i,T) \,,
\ee
with $d_i$ specifying the degeneracy (spin, isospin) of species $i$ and $m_i$ its mass. The sum runs over all species. For simplicity, we assume for the moment $\mu_B=0$. Here
\be \label{23}
\phi(m_i,T) = \int d^3p ~\exp\{\sqrt{p^2 + m_i^2}/T\}
\sim \exp-(m_i/T) \,,
\ee
is the Boltzmann factor for species $i$, so that the ratio of the production rates, $N_i$ and $N_j$, for hadrons of species $i$ and $j$ is
given by
\be
{N_i\over N_j}= {d_i \gamma_s^{\nu_i}\phi(m_i,T)
\over d_j \gamma_s^{\nu_j}\phi(m_j,T)} \,,
\ee
where $\nu_i=0,1,2,3$ specifies the number of strange quarks in species $i$.

Both temperature $T$ and strangeness suppression factor $\gamma_s$ have been measured, at various collision energies, and for different collision configurations. The resulting temperature of the emerging resonance gas is
found to have a universal value
\be
T_c \simeq 160 \pm 10 {\rm MeV} \,,
\ee
for all (high) collision energies, where $\mu_B \simeq 0$, and all collision configurations, including hadron production in $e^+e^-$ annihilation.

Moreover, in heavy ion collisions at lower energy, the finite baryon density has a crucial role and the dynamics is dominated by Fermi statistics and baryon repulsion. In the $T - \mu_B$ plane, the dependence of the hadronization temperature on
$\mu_B$ defines the chemical ``freeze-out'' curve, which can be described by specific, but poorly understood (but see next section), criteria \cite{magas,cley1,PBM3,cley3,taw}.

Indeed, a fixed ratio between the entropy density, $s$, and the hadronization temperature, $s/T^3\simeq 7$, or the average energy per particle, $< E > / N \simeq 1.08$GeV reproduce the curve in the $T - \mu_B$ plane as shown in Fig.1, where the percolation model
result \cite{magas} is also plotted.

\begin{figure}
\centerline{\includegraphics[width=0.7\textwidth]{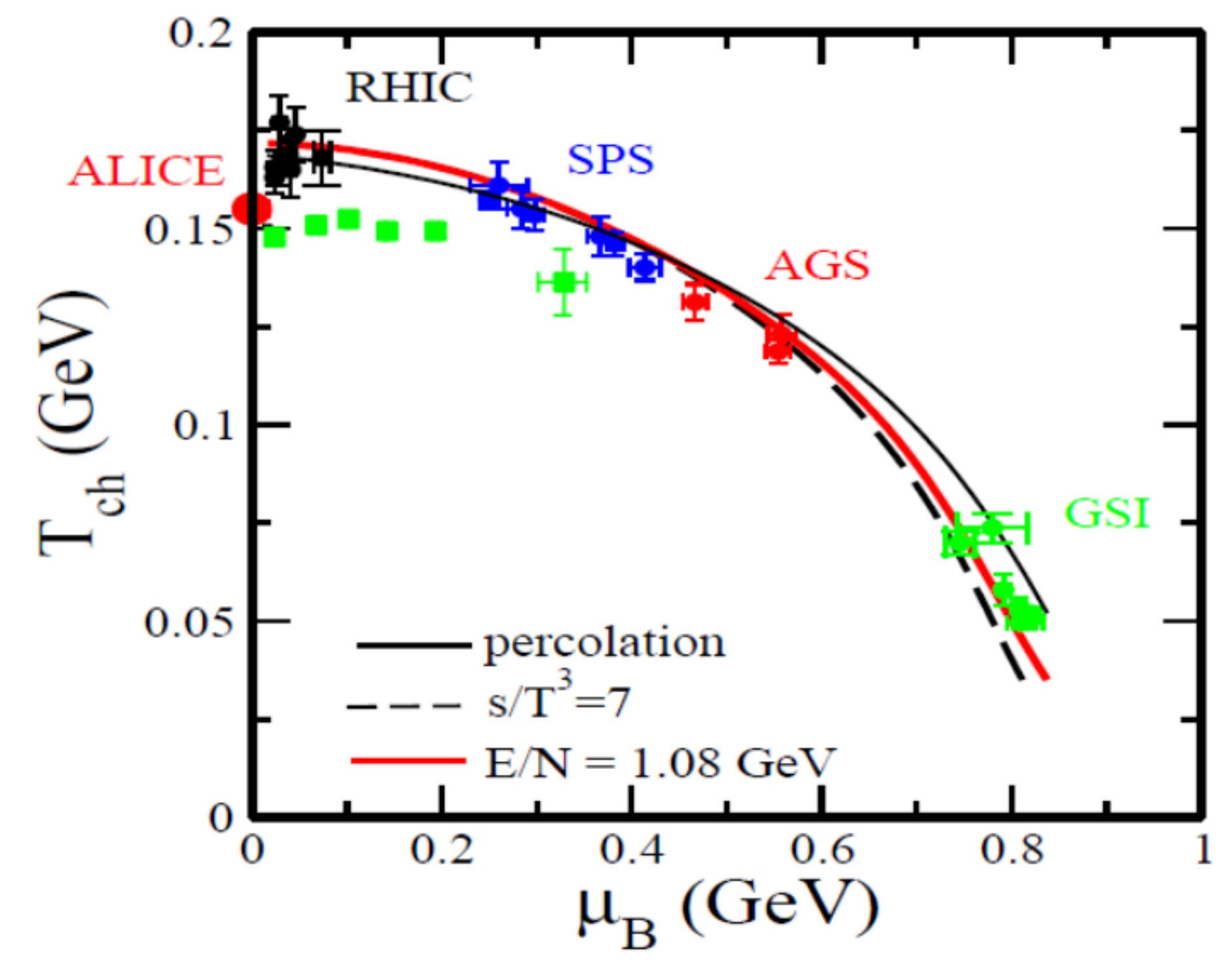}}
\caption{Freeze-out curve in the statistical hadronization model compared with the criteria discussed in the text. The green squares without error bars are the QCD lattice simulation data.}
\end{figure}

The agreement of the SHM with data on the abundances of different hadronic species, from $e^+ e^-$ annihilation till heavy ion collisions, is something puzzling. While in heavy ion collisions it is possible to expect the emergence of statistical distributions as a result of intense re–interactions, this seems very implausible in $e^+e^-$ annihilation at high energies, because there the density of the produced hadrons is small.

Moreover, in $e^+e^-$, the jet structure, the angular distributions of the produced hadrons, and the inter–jet correlations, point to the all–important role of QCD dynamics of gluon radiation. Thus the ``phase space dominance'' cannot be invoked. Indeed, in all high energy collisions, for $\sqrt(s) \ge 20$GeV, the hadronization temperature is essentially constant and independent from the initial configurations.

The previous aspects call for some \textit{universal} mechanism, at the root of hadron production, which has to be related to the way the QCD vacuum responds to color fields.

\subsection{Thermal component in the transverse momentum spectra in high energy hadronic processes}\label{TransverseMomentum}

The transverse momentum, $p_T$, spectra of hadrons produced in high energy collisions can be decomposed into two components: the exponential, or ``soft'', component and the power, or ``hard'', component.
Their relative strength, in deep inelastic scattering (DIS), depends drastically upon the global structure of the event. Namely, the exponential component is absent in the diﬀractive events characterized by a rapidity gap \cite{dima4,dimanew}.

The hard component is well understood, as resulting from the high-momentum transfer scattering of quarks and gluons and their subsequent fragmentation. The ``soft'' component is ubiquitous, in high energy collisions, and appears as a thermal spectrum. While in nuclear collisions, given the high number of participants involved, one may expect thermalization to take place, it is hard to believe that this might occur in processes as DIS or $e^+ e^-$ annihilation.

In \cite{Byl} it has been found that the following parametrization well describes the hadron transverse momentum distribution, both in hadronic collisions and in deep-inelastic scattering
\be
\label{eq:exppl}
\frac{d\sigma}{p_T d p_T} = A_{therm} \, e^{- m_{T}/T_{th}} +
A_{hard} \, \left(1+\frac{m_T^2}{n \, T_{th}^{2}}\right)^{-n} \,.
\ee
This clearly defines the soft/thermal component and the hard component parameterized by  $T_{th}$. Here $m_T = \sqrt{m^2 + p^2_T}$.

\subsection{Self-organization, self-similarity and the hadronic spectrum }\label{SelfOrganization}

The typical illustration of self-organized criticality (SOC), proposed in the pioneering work \cite{bak}, is the avalanche dynamics of sandpiles. There the number, $N(s)$, of avalanches of size $s$, observed over a long period, is found to vary as a power of $s$, $N(s) = \alpha s^{-p}$.
This means that the phenomenon is scale-free, so the same structure is found, again and again, at all scales. This phenomenon is often referred to as self-similarity: the system resembles itself at all scales.

Another example of self-similarity is found when partitioning naturals. Given a natural number, $N \in \mathbb{N}$, we can \textit{decompose} it, in mathematical jargon, into the naturals whose sum gives $N = \sum_i N_i$, with no distinction of the order the $N_i$s enter the sum. E.g., 3 = 2+1 and 3 = 1+2, would count the same as a decomposition of 3. On the other hand, we also have \textit{compositions} of $N$, that are decompositions of $N$ in which the \textit{order} of the terms matters. In the following, according to the abuse of language of the physics literature, we shall call the decompositions ``unordered partitions of the integer'' (UPIs), and the compositions ``ordered partitions of the integer'' (OPIs).

The number of OPIs of $N$, say it $O(N)$, can be easily computed to be
\be \label{selfsimOPI}
O(N) = 2^{N-1} \,.
\ee
In other words, the self-similarity pattern can be phrased as ``large integers consist of smaller integers, which in turn consist of still smaller integers, and so on...''.

Starting with the integer $N$, we would now like to know the number $n(N,k)$ that specifies how often a given integer $k$ occurs in the set of all OPIs of $N$. E.g., considering again $N=3$, we have $n(3,3)=1$, $n(3,2)=2$ and $n(3,1)=5$. To apply the formalism of
SOC, we associate a weight $s(k)$ to each integer. The natural choice is $s(k) = O(k) = 2^{k-1}$ and the number $n(N,k)$ we are looking for, in a scale-free scenario, is then given by
\be
n(N,k) = \alpha(N) [s(k)]^{-p} \,.
\label{p3}
\ee
For small values of $N$, $n(N,k)$ is readily obtained explicitly and one finds that the critical exponent is $p \simeq 1.26$.

The previous example is immediately reminiscent of the statistical bootstrap model of Hagedorn \cite{hagedorn,frautschi,nahm,hagedorn3}. There we have ``fireballs composed of fireballs, which
in turn are composed of smaller fireballs, and so on''. Indeed, its general pattern has been shown to be due to an underlying structure, related to the OPIs \cite{blan}.

More precisely, Hagedorn's bootstrap approach proposes that a hadronic colorless state, with overall mass $m$, can be partitioned into structurally similar colorless states. Then, those component colorless states can be partitioned, into structurally similar colorless states, and so on. If the states were at rest, the situation would be identical to OPI just discussed. Since the constituent fireballs, though, have an intrinsic motion, the number of states, $\rho(m)$, corresponding to a given mass $m$, is determined by the bootstrap equation which can be asymptotically solved \cite{nahm}. This gives
$\rho(m) \sim m^{-a}$ exp${(m/T_H)}$, and $T_H$ is solution of
\be
\left({2\over 3 \pi}\right)\left(T_H \over m_0\right)
K_2(m_0/T_H) = 2 \ln 2 - 1,
\label{bs}
\ee
with $m_0$ denoting the lowest possible mass and $K_2(x)$ denoting a Hankel function of pure imaginary argument. For $m_0 = m_{\pi} \simeq 130$ Mev, this leads to the Hagedorn temperature
\be
T_H \simeq 150 {\rm MeV} \,,
\ee
that is, approximately, the critical hadronization temperature found in statistical QCD. The cited solution gave $a=3$, but other exponents could also been considered.

The previous expression of $\rho(m)$ is an asymptotic solution of the bootstrap equation, which diverges for $m\to 0$, hence it cannot hold for small masses. Using for $\rho(m)$ a result similar to the one obtained in the dual resonance model,
Hagedorn proposed
\be
\rho(m) = {\rm const.}(1+ (m/\mu_0))^{-a} \exp(m/T_H)
\label{exp1} \,,
\ee
where $\mu_0 \simeq 1 - 2$ GeV is a normalization constant.

At this point we should emphasize that the form of $\rho(m)$ is entirely due to the self-organized nature of system. That is in no way a result of a thermal behavior. We have expressed the slope coefficient of $m$  in terms of the Hagedorn
``temperature'' only because we have in mind the analog gravity scenarios that will be soon discussed, but, by itself, that coefficient is exclusively of combinatorial origin.


\section{Analog gravity interpretation of the SHM and the QCD Hawking-Unruh radiation}\label{SecStatHadrModel}


As mentioned in the Introduction, the phenomenology of quark confinement can be seen as a the effect of a Rindler force due to the string tension, $\sigma$. Let us now describe this phenomenon in more detail.

\begin{figure}[h]
\centerline{\includegraphics[width=0.6\textwidth]{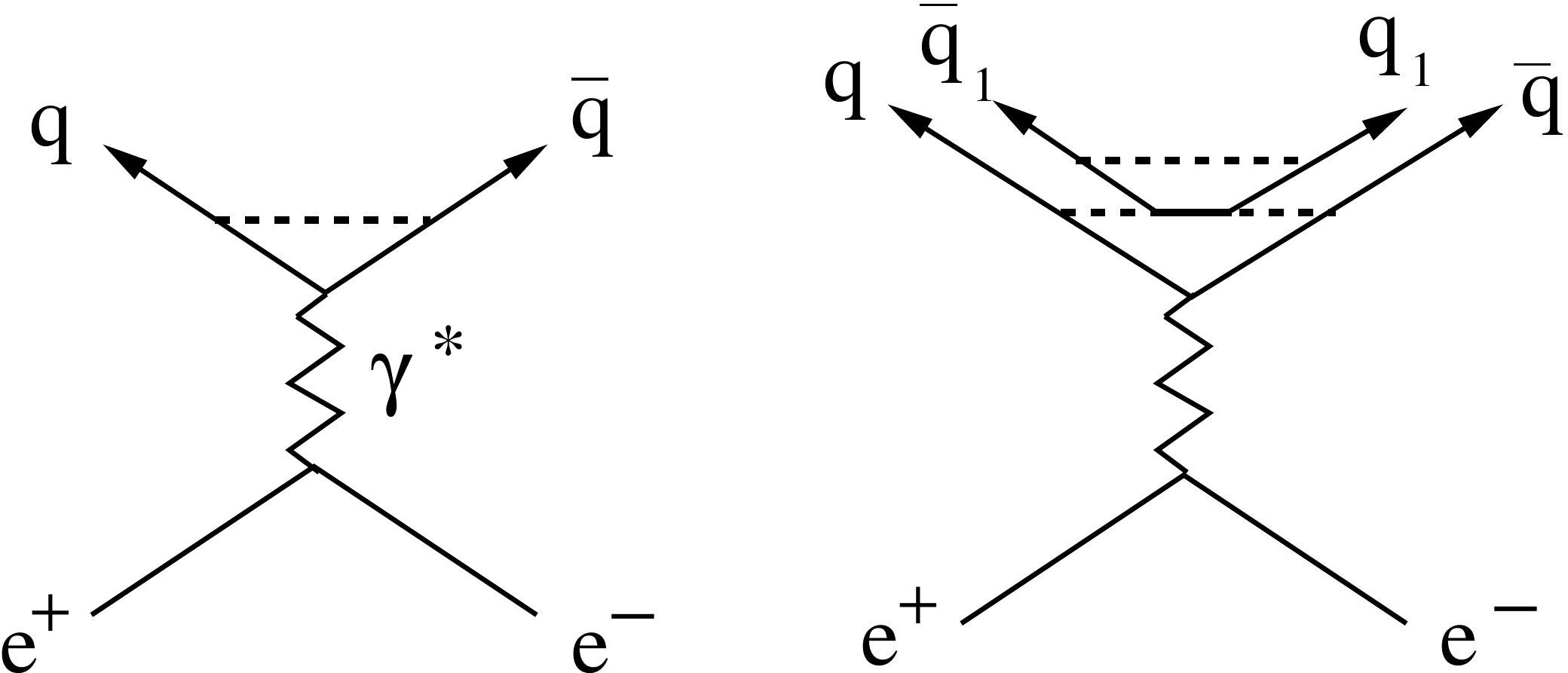}}
\caption{Quark formation in $e^+ e^-$ annihilation}
\label{anni}
\end{figure}

We have recalled in Section \ref{ReviewUnruhHawking} that the basic mechanism of Unruh radiation is tunnelling through the confining event horizon. This is most simply illustrated by hadron production through
$e^+ e^-$ annihilation into a $q$ pair, see Fig. \ref{anni}.

The first quark-antiquark pair, $q \bar{q}$, initially tries to separate. The attempt stops when both quarks hit the confinement horizon, i.e., when they both reach the end of the binding string, where their separation is $R$. At that point, the attempt to separate can only continue
if a further quark-antiquark system is excited from the vacuum. Although the new pair, $q_1 \bar{q}_1$, is at rest, in the overall center of mass system, each of its constituents has a transverse momentum $k_T$, determined by the uncertainty relation in terms of the transverse dimension of the string
flux tube. String theory \cite{Lue} gives for the basic thickness
\be
r_T=\sqrt{2/\pi \sigma} \,,
\label{1}
\ee
leading to
\be
k_T=\sqrt{\pi \sigma/2} \,.
\label{2}
\ee
The maximum separation distance $R$ is thus specified by
\be
\sigma R = 2 \sqrt{m_q^2 + k_T^2} = 2k_T \,,
\label{3}
\ee
where we have taken $m_q=0$ for the quark mass. From this we obtain
\be
R = \sqrt{2 \pi / \sigma} \,,
\label{4}
\ee
as the string breaking distance. The departing quark $q$ now pulls the
newly formed $\bar q_1$ along, giving it an
acceleration
\cite{Castorina2007}
\be
a = \sqrt{2 \pi \sigma} \,.
\label{5}
\ee
The $q_1\bar q_1$ pair eventually suffers the same fate as the $q$ pair:
it is separated up to its confinement horizon, where it again excites a new
pair, which is now emitted as Unruh radiation of temperature
\be
T_h = {a / 2\pi} = \sqrt{\sigma/2 \pi} \,,
\label{6}
\ee
that is also the hadronization temperature, as we shall see in a moment. This process is sequentially repeated until the energy of the initial ``driving'' quarks $q$ and $\bar q$ is exhausted.

\begin{figure}[h]
\centerline{\includegraphics[width=0.6\textwidth]{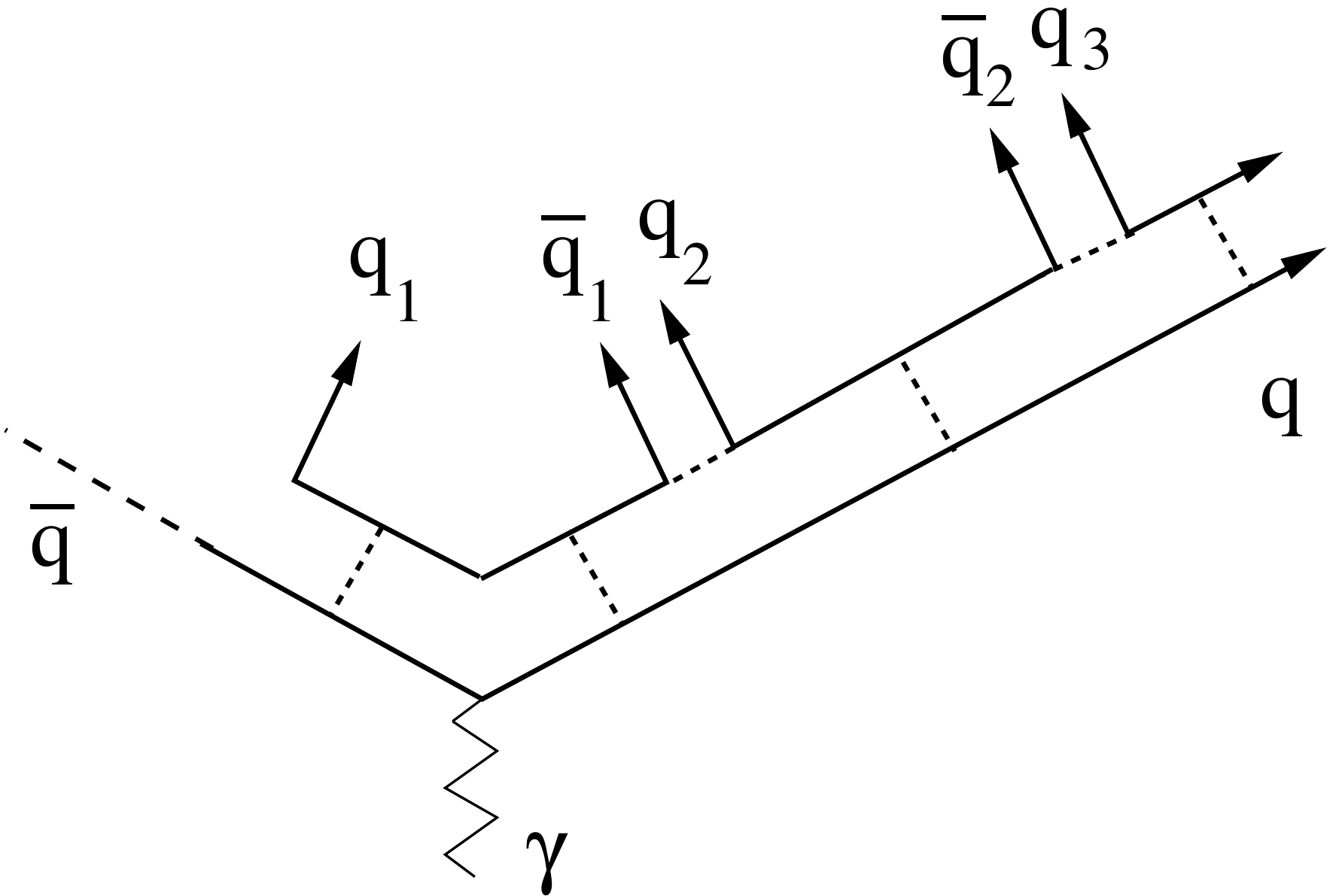}}
\caption{Sequential quark formation in $e^+ e^-$ annihilation}
\label{anni2}
\end{figure}

The case of the $e^+ e^-$ annihilation corresponds to baryochemical potential, $\mu_B=0$. Here one finds the average value $\sigma \simeq 0.19 \pm 0.03 $ GeV$^2$, see, e.g., \cite{sigma1}, which with Equation (\ref{6})
then leads to
\be
T_h(\mu_B=0)  = \sqrt{\sigma/2 \pi} \simeq 175 \pm 15~{\rm MeV} \,.
\label{7}
\ee
for the freeze-out temperature at $\mu_B=0$.

The fundamental mechanism in the Unruh scenario is quark (de)acceleration leading to string breaking with the resulting pair production, as specified by Equation
(\ref{3}). As long as we assume a vanishing quark mass, the only dimensional parameter in the entire formalism then is the string tension $\sigma$.

The Unruh hadronization temperature is therefore ``universal'', and this explains the observation of thermal hadron production in high energy
collisions in $e^+ e^-$ and $pp$ interactions also. In this respect the emitted hadrons are ``born in equilibrium'' \cite{hagedorn,hagedornBorn}.


\medskip

The previous analysis shows that the hadronization temperature corresponds to the Unruh temperature related with the string breaking in high energy collisions, where $\mu_B \simeq 0$.

As discussed, the dependence of the hadronization temperature on $\mu_B$ defines the chemical ``freeze-out'' curve, which turns out to be in agreement (see Fig.1.) with a
fixed ratio between the entropy density, $s$, and the hadronization temperature, $s/T^3\simeq 7$ and/or the average energy per particle, $< E > /N \simeq 1.08$ GeV and/or
$n \simeq 0.12$ fm$^{-3}$, where $n$ is the number density.

Although the Unruh mechanism and the string breaking provide a theoretical basis for the production of newly formed hadrons in high energy collisions, they do not address the role of the nucleons already
present in the initial state of the heavy ion collisions. However, the corresponding hadron formation gives clear meaning to the figures that characterize the whole freeze-out curve.

Indeed, as discussed in \cite{caiosa1}, the energy of the pair produced by string breaking, i.e., of the newly formed hadron is given by (cf Equations \ (\ref{2}) and (\ref{3}))
\be
E_h = \sigma R = \sqrt{2 \pi \sigma} \,.
\label{8}
\ee
In the central rapidity region of high energy collisions, one has
$\mu_B \simeq 0$, so that $E_h$ is in fact the average energy
$\langle E \rangle$ per hadron, with an average number $\langle N
\rangle$ of newly produced hadrons.
Hence one obtains
\be
{\langle E \rangle \over \langle N \rangle
} = \sqrt{2 \pi \sigma} \simeq 1.09 \pm 0.08 \, {\rm GeV} \,,
\label{9}
\ee
in accordance with the phenomenological fit obtained from the species abundances in high energy collisions \cite{cley1,cley3}.

Next, we turn to the number density.  For a single string-breaking the number density is given by
\be
n_{sb} \simeq \frac{1}{4\pi R^3/3} \,,
\ee
where $R$ is the string breaking distance, which turns out to be $R = 1/T_h$, for massless quarks. For $T_h \simeq 160$ MeV, consistently with our previous evaluation, one obtains $n_{sb} \simeq 0.129$
fm$^{-3}$.

Let us now consider the entropy. Since the event horizon is caused by color confinement, such an entropy is necessarily an entanglement entropy of quantum field modes on both sides of the horizon (recall that here we have no real gravitational degrees of freedom). Its general
form is \cite{tera,alfgae}
\be
\label{entropymatter}
S_{\rm ent} = \alpha \frac{A}{r^2} \,,
\ee
where $A$ is the area of the event horizon, $r$ the scale of the characteristic quantum fluctuations and $\alpha$ an undetermined numerical constant, that might as well be infinite. This expression shares the holographic
structure (holography of entanglement entropy is a quite general result, see \cite{sre,Solodukhin}) with the Bekenstein-Hawking entropy \cite{bek,haw1} for a BH given in (\ref{SBekHaw})
\[
S_{\rm BH} = {1 \over 4} ~ {A \over \ell_P^2} \,.
\]

\bigskip

A relation similar to (\ref{SBekHaw}) also holds in the case of an accelerated observer \cite{laflamme}. Here we take it to be valid also in our case, where gravity is not involved and the entire entropy must be of the entanglement type.
The scale of the characteristic quantum fluctuations is now given by the transverse string thickness in Equation (\ref{1}), rather than the Planck length, $\ell_P$, of gravitational phenomena. One obtains
\be
S_h = {1\over 4}~ {A_h \over r_T^2} = {1\over 4}~ {4 \pi R^2 \over r_T^2} \,,
\label{11}
\ee
for the entropy in hadron production. The parameter $R$ is given by Equation (\ref{4}) and inserting these expressions into Equation (\ref{11})
gives for the entropy associated to hadron production
\be
S_h = \pi^3 \,,
\label{S}
\ee
and the entropy {\sl density}, $s = S_h/V$ (here $V= 4/3 \pi R^3$), divided by $T^3$ turns out to be
\be
{s  \over  T^3} = {S_h \over (4 \pi/3) R^3 T^3} =
{3 \pi^2 \over 4} \simeq 7.4 \,,
\label{entrop}
\ee
as freeze-out condition in terms of $s(T)$ and $T$. This result is in accordance with the value obtained for $s/T^3$ from species abundance analyses in terms of the ideal resonance gas model \cite{cley3,taw}.
Moreover, within this picture one can show \cite{caio1} that QCD entropy, evaluated by lattice simulations in the region $T_c < T < 1.3T_c$, is in reasonable agreement with a melting color event horizon.

The analogy between the freeze-out temperature as a function of $\mu_B$ and the Hawking temperature for charged BH has been discussed in \cite{Castorina2007} and another interesting aspect is that it can be translated in the temperature dependence on the collision energy $\sqrt{s}$, by considering $\mu_B(\sqrt s)$ \cite{pi3}.

Since the Unruh temperature triggers the search for the gravitational BH that in its near-horizon approximation better simulates the hadronization phenomenon, one can study  which BH behind that Rindler horizon could reproduce the experimental behavior of $T(\sqrt{s})$. Although the complete hadronization process is in 4D space-time, the hadronic Rindler spacetime should be better consider as  the near-horizon approximation of the effective two-dimensional (2D) BH analog for the following two reasons
\begin{itemize}
  \item New particle creation is effectively 2D because it can be described in terms of the evolution in time of the hadronic strings, that are one dimensional objects \cite{grumi3}.
  \item The near horizon field dynamics is effectively 2D \cite{siddhartha,parwilc}.
\end{itemize}
Provided certain natural assumptions hold, it has been shown \cite{grumi} that the, so called, exact string BH in 2D dilaton gravity \cite{grumi3} turns out to be the best candidate, as it fits the available data on $T(\sqrt{s}$), and that its limiting case, the Witten BH, is the unique candidate to explain the constant $T$ for all elementary scattering processes at large energy.


\bigskip

To close this Section, we turn now to the strange quark mass and the interpretation {\it alla} Unruh of the strangeness enhancement.


At the beginning of this Section we have illustrated how the thermal hadron production process is a Hawking-Unruh mechanism. In doing so we have neglected the effects of the quark mass. If one includes them, the expression one obtains for acceleration is
\be
a_q = {\sigma \over w_q} =
{\sigma \over \sqrt{m_q^2 + k_q^2}},
\ee
where $w_q =\sqrt{m_q^2 + k_q^2}$ is the effective mass of the produced quark, with $m_q$ the bare quark mass and $k_q$ the quark momentum inside the hadronic system
$q_1\bar q_1$ or $q_2\bar q_2$ (see Fig. 3). Since the string breaks \cite{Castorina2007} when it reaches a separation distance
\be
x_q \simeq {2\over \sigma} \sqrt{m^2_q + \frac{\pi \sigma}{2}},
\ee
the uncertainty relation gives us with $k_q \simeq 1/x_q$
\be
w_q  = \sqrt{m_q^2 + [\sigma^2/(4m_q^2 + 2\pi \sigma)]} \,,
\ee
for the effective mass of the quark. The resulting Unruh temperature depends now from the quark-mass, and it is thus given by
\be
T(qq) \simeq {\sigma \over 2\pi w_q} \,.
\label{Tq}
\ee
Note that here it is assumed that the quark masses for $q_1$ and $q_2$ are equal. For $m_q \simeq 0$, Equation (\ref{Tq}) reduces to $T(00) \simeq \sqrt{\sigma / 2\pi}$, as obtained in Equation (\ref{6}).

If the produced hadron ${\bar q}_1 q_2$ consists of quarks of different masses, the resulting temperature has to be calculated as an average of the different accelerations involved. For one massless quark ($m_q \simeq 0$) and one of strange quark of mass $m_s$, the average
acceleration becomes
\be\label{avgaccel}
\bar a_{0 s} = {w_0 a_0 + w_s a_s \over w_0 + w_s} =
{2\sigma \over w_0 + w_s} \,.
\ee
From this, the Unruh temperature of a strange meson is given by
$T(0s) \simeq {\sigma / \pi (w_0 + w_s)}$ with $w_0 \simeq \sqrt{1/2\pi\sigma}$ and $w_s$ given by Equation (4) with $m_q=m_s$. Similarly, we obtain
$T(ss) \simeq {\sigma / 2 \pi w_s}$
for the temperature of a meson consisting of a strange quark-antiquark pair
($\phi$).

The scheme is readily generalized to baryons. The production
pattern leads to an
average of the accelerations of the quarks involved \cite{firenze1}. We thus have
$T(000) = T(0) \simeq {\sigma / 2\pi w_0}$for nucleons, $T(00s) \simeq {3 \sigma / 2\pi(2w_0 + w_s)}$
for $\Lambda$ and $\Sigma$ production,
$T(0ss) \simeq {3 \sigma / 2\pi(w_0 + 2w_s)}$ for $\Xi$ production, and
$T(sss) = T(ss) \simeq {\sigma / 2\pi w_s}$
for that of $\Omega$s.

We thus obtain a resonance gas picture with five different hadronization
temperatures, as specified by the strangeness content of the hadron in
question: $T(00)=T(000),~T(0s),~T(ss)=T(sss),~T(00s)$ and $T(0ss)$.

In other terms, the event horizon of colour confinement leads to thermal behaviour, but the resulting temperature depends on the strange quark
content of the produced hadrons, causing a deviation from full equilibrium and hence a suppression of strange particle production, without the introduction of the $\gamma_s$ parameter. The resulting formalism has been applied
to multihadron production in $e^+ e^-$ annihilation  over a wide range of energies to make a comprehensive analysis of the data, in the conventional (i.e. with $\gamma_s$) SHM and its modified Hawking-Unruh formulation \cite{firenze1,firenze2}.
The modified SHM , with the different Unruh temperature gives a better fit with respect to the standard SHM formulation.

\begin{figure}[h]
\begin{minipage}[t]{7.5cm}
\includegraphics[width=0.9\textwidth]{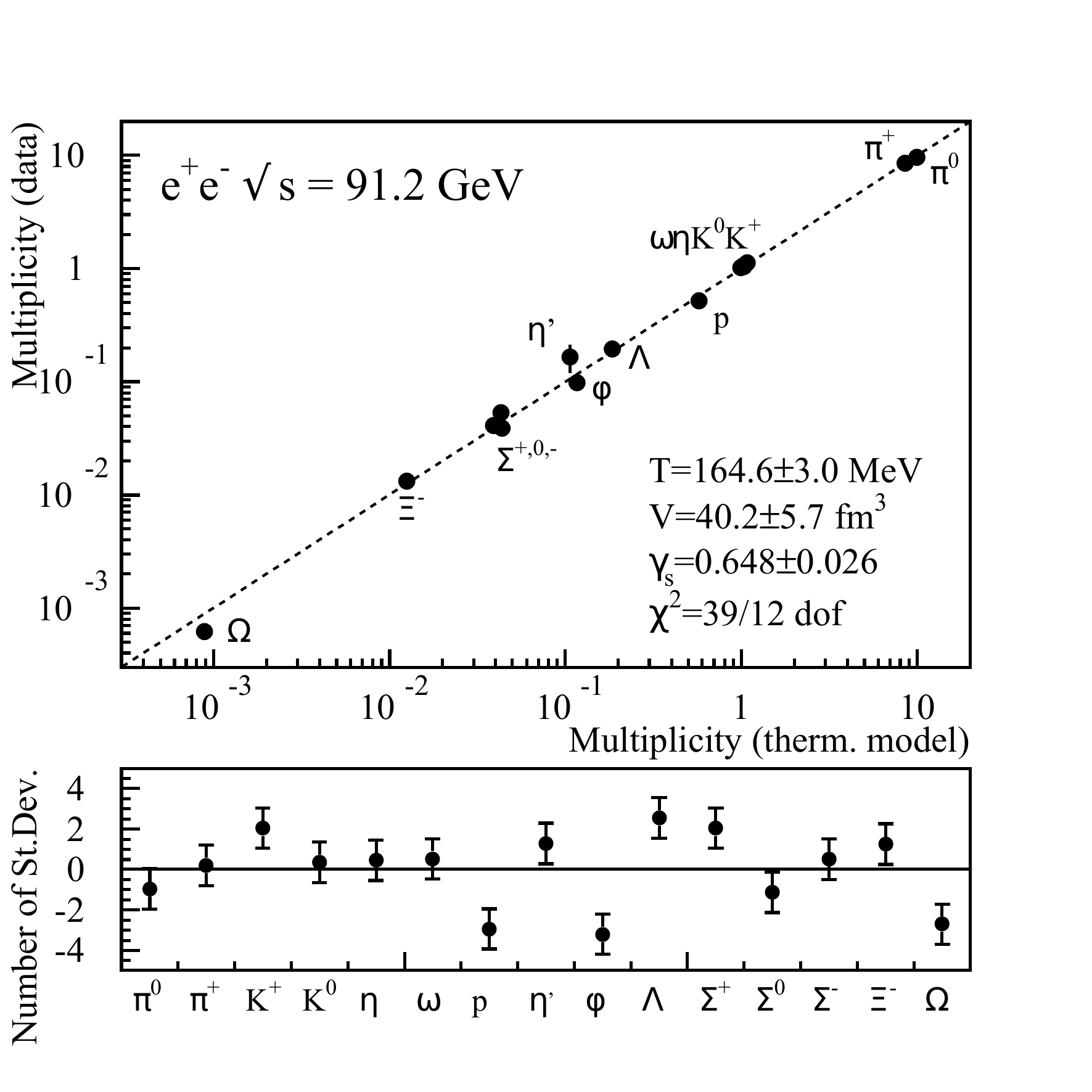}
\end{minipage}
\begin{minipage}[t]{7.5cm}
\includegraphics[width=0.9\textwidth]{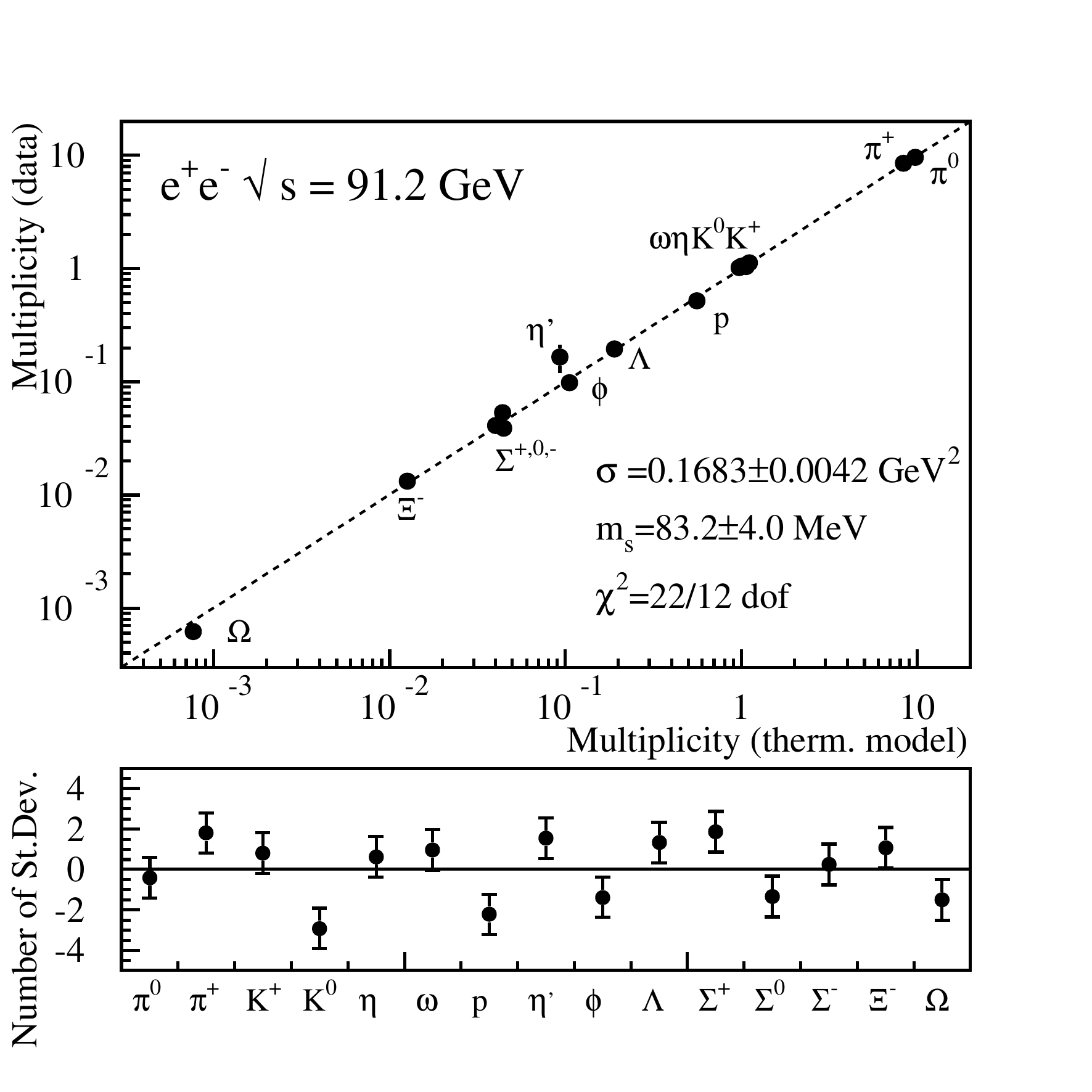}
\end{minipage}
\caption{Comparison between measured and fitted multiplicities of long-lived
hadronic species in $e^+ e^-$~collisions at $\sqrt s = 91.25$ GeV. Left:
statistical hadronization model with one temperature. Right: Hawking-Unruh
radiation model. See \cite{firenze1}}
\label{ee91plot}
\end{figure}

It should be noted that  in the Hawking-Unruh formulation the number of free parameters of the model does not increase since all the previous temperatures are completely determined by the string tension and the strange quark mass. Apart from possible variations of the quantities of $\sigma$ and $m_s$, the description is thus parameter-free.

It is seen that, in all cases, the temperature for a hadron carrying nonzero strangeness is lower than that of non-strange hadrons and this leads to an overall strangeness suppression in elementary collisions, in good agreement with data, without the introduction of the ad-hoc parameter $\gamma_s$.
 Fig.4. reports the comparison between the SHM with one temperature and $\gamma_s$ and the Hawking-Unruh-inspired approach.

On the other hand, in nucleus-nucleus (AA, ``large systems'') collisions at $\sqrt s \geq 15$ GeV the so called strangeness enhancement with respect to $e^+e^-$ and hadronic scatterings (the ``small'' systems) has been observed, which in the standard SHM is described by the condition $\gamma_s=1$ in AA with respect to $\gamma_s \simeq 0.5-0.6$ in small systems. Moreover the same enhancement has been detected in proton-proton collisions at large energy and in large multiplicity events \cite{Adam2017}.

The translation {\it alla} Unruh of the strangeness enhancement requires that the different temperature for various hadronic strangeness  content has to disappear in those cases. Indeed, $T(00),T(0s),...$ derive from the breaking of a single string with the corresponding average acceleration and Unruh temperatures. On the other hand, as shown in ref.\cite{nostromartin}, the universality among small and large systems is directly related with the initial parton density in the transverse plane.

If the initial setting are different but the collision energy and the large multiplicity cut produce initial states with similar entropy-density (i.e parton density in the transverse plane) the hadron production and other coarse-grain dynamical signatures are the same \cite{nostromartin}. Therefore, for large parton density, there is a strong string overlap as depicted in Fig.5.

Let us outline, in a simplified model, the mechanism which washes out the strangeness dependence of the Unruh temperature when, in a causally connected region, the parton density in the transverse plane is large.

Assume two species only: one scalar meson and one electrically neutral meson, that is, ``pions'' with mass $m_\pi$, and ``kaons'' with mass $m_k$ and strangeness $s=1$.

Let us consider a high density system of quarks and antiquarks in a causally connected region for high energy and high multiplicity events. Generalizing Equation (\ref{avgaccel}) the average acceleration is given by
\be
\bar a = \frac{N_l w_0 a_0 + N_s w_s a_s}{N_l w_0 + N_s w_s} \,,
\ee
where $N_l>>1$,  $N_s>>1$ are respectively the number of light quarks and of strange quarks.

\begin{figure}[h]
\begin{minipage}[t]{6.5cm}
\includegraphics[width=0.9\textwidth]{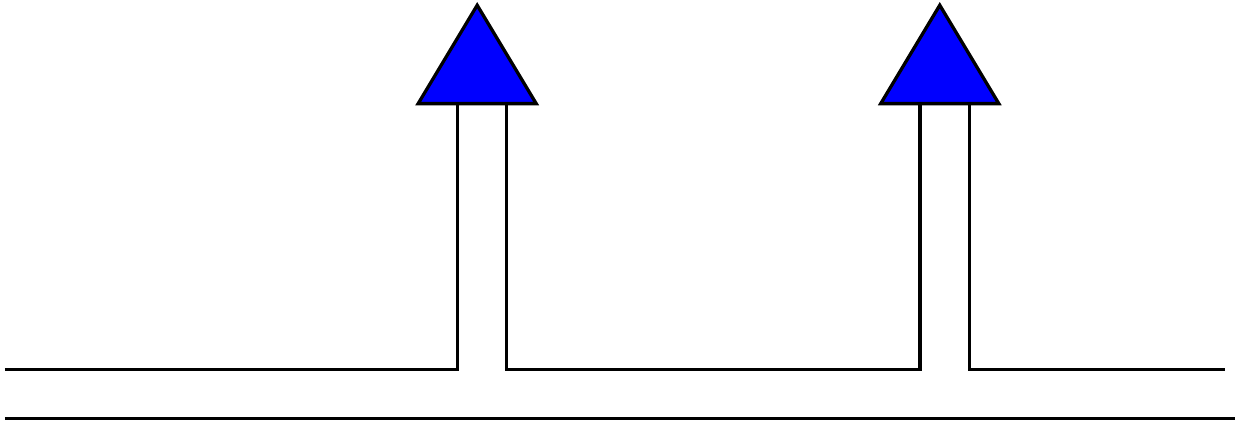}
\end{minipage}
\begin{minipage}[t]{6.5cm}
\includegraphics[width=0.9\textwidth]{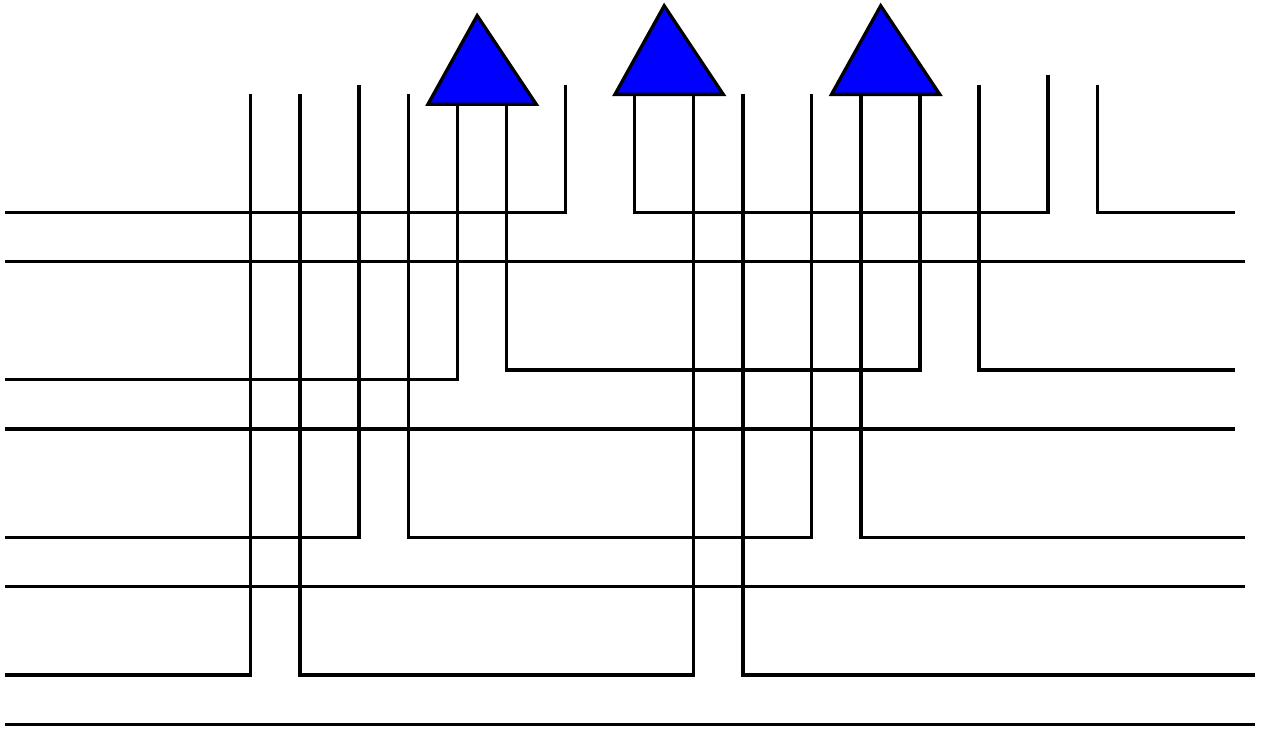}
\end{minipage}
\caption{Left: Hadron production {\it alla} Unruh by a sequence of independent single string breakings.
Right: Hadron production due to the overlap of different color event horizon for large parton density.}
\label{recomb}
\end{figure}

By assuming $N_l >> N_s$, after a simple algebra, the average temperature, $\bar T = \bar a/2\pi$,  turns out to be
\be
\bar T = T(00)\left[ 1 - \frac{N_s}{N_l} \frac{w_0 + w_s}{w_0} \left(1 - \frac{T(0s)}{T(00)}\right)\right] + O[(N_s/N_l)^2] \,,
\ee
Now in our ``world of pions and kaons'' one has $N_l = 2 N_\pi + N_k$ and $N_s=N_k$ and therefore
\be \label{T00}
\bar T = T(00) \left[ 1 - \frac{N_k}{2 N_\pi} \frac{w_0 + w_s}{w_0} \left(1 - \frac{T(0s)}{T(00)}\right)\right] + O[(N_k/N_\pi)^2] \,.
\ee
On the other hand, in the Hawking-Unruh based statistical calculation the  kaon-pion ratio, $N_k/N_\pi$, depends on the equilibrium (average) temperature $\bar T$, that is
\be \label{NkNpi}
N_k/N_\pi = \frac{m_k^2}{m_\pi^2} \, \frac{K_2(m_k/\bar T)}{K_2(m_\pi/\bar T)} \,,
\ee
where $K_2(x)$ denotes a Hankel function of pure imaginary argument. Therefore, one has to determine the temperature $\bar T$ by self-consistency of Equation (\ref{T00}) with Equation (\ref{NkNpi}). This condition implies
\be
\frac{2 \, [1 - \bar T/T(00)]w_0}{[1-T(0s)/T(00)](w_s+w_0)} =  \frac{m_k^2}{m_\pi^2} \, \frac{K_2(m_k/\bar T)}{K_2(m_\pi/\bar T)} \,,
\ee
that can be solved numerically. For $\sigma=0.17$ Gev$^2$, $m_s=0.083$GeV (see Fig.4), the solution gives $\bar T/T(00) \simeq 0.97$.

In other words, this toy model shows that the non-equilibrium condition, with species-dependent temperatures, converges to an equilibrated system, with the average temperature, $\bar T \simeq T(00)$, for large parton density in a causally connected region.

\section{Analog gravity interpretation of the origin of thermal component in the transverse momentum spectra}\label{SecTransverseMomentum}

The strength of the chromoelectric field, in a single string breaking, is determined by the string tension and it describes the yields of the different hadronic species. However, to discuss the transverse momentum spectra of the produced hadrons (see Subsection \ref{TransverseMomentum}) one has to take into account the increasing number of gluons in the wave functions of the colliding hadrons. This can be done by the parton saturation \cite{Gribov:1984tu}, or color glass condensate \cite{McLerran:1993ni,McLerran:1994} picture. In this approach the density of partons, in the transverse plane, is parameterized by the saturation momentum $Q_s(s, \eta)$ that depends on the c.m.s. collision energy squared $s$ and (pseudo-)rapidity $\eta$.

The temperature of the radiation from the resulting Rindler event horizon is thus given by \cite{KHARZEEV2005316}
\be\label{unruh}
T_U = T_{th} = c\ \frac{Q_s}{2 \pi} ,
\ee
where $c$ is a constant \cite{dima5}. $T_{th}$ is related to the deceleration of partons in the transverse plane, moreover $Q_s = T$ in the parametrization of the hard component in Equation (\ref{23}) \cite{dima4}. Therefore one predicts a proportionality between the $T_{th}$ and $T$, which has been verified \cite{dima4,dimanew}.

The established proportionality of the parameters describing the thermal and hard components of the transverse momentum spectra supports the theoretical picture in which the soft hadron production is a
consequence of the quantum evaporation from the event horizon formed by deceleration in longitudinal color ﬁelds. The absence of the thermal component in diﬀractive interactions lend further support to this interpretation.

\section{Analog gravity interpretation of the partitions of integers for the BHs self-similarity}\label{SecSelfOrganization}

The celebrated \textit{self-similarity} at work in the hadronic spectrum, recalled in Subsection \ref{SelfOrganization}, is typical of many physical setups that enjoy scale invariance, like fractals, phase transitions at the critical point, etc \cite{zinnjustin}. Among those, BH self-similarity \cite{Harms:1992nb,Harms:1992jt,Huang:2000kga} is surely one of the most interesting, if one wants to probe fundamental ideas of QG.

Some aspects of BH self-similarity are understood, if one recalls that the Hawking temperature, $T_H$, of a Planck-sized BH ($T_H \approx l^{-1}_P$, where $l_P$ is the Planck length) could be viewed as the Hagedorn temperature in string theory \cite{veneziano1,susskind1,susskindBOOK}. At that temperature, BH evaporation stops and a phase transition is expected to occur, in analogy to what happens at the phase transition between the hadrons and the quark-gluon-plasma phase \cite{Cabibbo:1975ig}. Nonetheless, to properly speak of self-similarity, one would really need to make sense of statements like ``large BHs could be viewed as formed by smaller BHs, formed in turn of even smaller BHs, and so on...''

In the work \cite{progress} some steps were moved in that direction, and a link was established, in simple terms, between the space of BH configurations and the OPI. This, in turn, shed a new light on BH self-similarity, in the plain terms of the statement quoted above. In what follows, let us comment on this.

First and foremost, the model we refer to is the so-called ``quantum BH'' of Mukhanov and Bekenstein, \cite{mukhanov,bekenstein2, bekenstein3}. In that approach the area of the BH's event horizon is quantized
\be
A  \ = \  \alpha \, N  \, l_P^2 \,,
\ee
where $N \in \mathbb{N}$ and the ``it from bit'' \cite{wheelerit} choice for the proportionality factor, $\alpha = 4 \ln 2$, allows for a two-level spin-$1/2$ system description, $\uparrow$ or $\downarrow$, per given Planck cell. With these, BH entropy, $S_{BH}$, can be written as
\be \label{ClassicalEntropy}
S_{BH} \ = \ \frac{A }{4 \, l_P^2} \ = \ N  \, \ln 2 \,,
\ee
which is the entropy of a quantum system living in a Hilbert (configuration) space of dimension dim$H = 2^N$, where each of the $2^N$ configurations has the same statistical weight, see, e.g., \cite{kiefer} for this and other approaches.

Thus, on the one hand, the number of OPIs of $N$, $O(N) = 2^{N - 1}$, whereas the number of configurations of the quantum BH is given by $C(N) = 2^N$. Therefore, if we want to relate the two ways of counting configurations, one needs to find a 2-to-1 map from the latter to the former.

In \cite{progress} this is achieved by distinguishing between BH configurations differing not only by how many spin are up and how many are down, as done in other approaches \cite{kiefer}, but also by the \textit{position} of the spin. The ``spin-flip map'', there introduced, does the job of halving the number of BH configurations in a consistent way, in order to associate spin states, on the one hand, with the OPI of $N$, on the other hand.

The 2-to-1 map works as follows: To any one given OPI of $N$ it associates the two BH states that are obtained one from the other when all the spins, that identify the given configuration, are flipped, $\uparrow \leftrightarrow \downarrow$. Then, the rule that relates a given \textit{pair} of BH (spin) configurations to a given OPI is (for details see \cite{progress}):

\textit{When a spin is next to an opposite spin, i.e., when $\uparrow$ is next to $\downarrow$ or when $\downarrow$ is next to $\uparrow$, in the OPI this corresponds to 1 + 1. E.g., $(\uparrow, \downarrow, \uparrow, ...)$ and the spin-flipped $(\downarrow, \uparrow, \downarrow, ...)$ both correspond in the OPI to the partition $1+1+ ...$. When the spin are likewise they contribute with an integer that is the sum of how many times the spin does not flip. E.g., $(\uparrow, \uparrow, \downarrow, ...)$ and $(\downarrow, \downarrow, \uparrow, ...)$ correspond in the OPI to the partition $2 + ...$.}

With these, one takes into account all possibilities, hence the wanted 2-to-1 map from the BH configurations to the OPI (the ``spin-flip map'') is obtained. Having established that, we want now to see how the self-similarity patterns of the OPI can be imported into the self-similarity of BHs.

To avoid overcounting some configurations or missing others, in \cite{progress} the authors construct an operation, $\hat{+}$, that allows to obtain the configuration space of the given BH only once, for any given partition. If we indicate with $\mathbf{N}$ such $2^N$-dimensional configuration space, and $N_1 + N_2 + \cdots = N$ is a given OPI of $N$, such an operation must give $\mathbf{N_1} \hat{+} \mathbf{N_2} \hat{+} \cdots = \mathbf{N}$. Doing so we would establish a one-to-one correspondence between the OPI of $N$, and the way to combine the subspaces of $\mathbf{N}$, corresponding to the OPI. We report here the actual definition of such an operation:

\textit{Take each partition of $N$, say $N_1 + N_2 = N$, and write the spin configuration space associated to the first number of the sum, $\mathbf{N_1}$. Then, take the tensor product of each of those representatives with all the spin configurations of $\mathbf{N_2}$, explicitly including all the spin-flipped configurations. The result of such operation, $\mathbf{N_1} \hat{+} \mathbf{N_2}$, are all the spin configurations of $\mathbf{N}$, with no redundant nor missed configuration. The operation gives the same result for each OPI of $N$, including those with more than two terms. For the latter, one must start from the first term on the left, act with the second as just described, and the result of this needs be acted upon with the next term, and so on, till the end.}

The trivial example is $\mathbf{N}=\mathbf{N}$, where no composition is performed. The first non-trivial operation is $\mathbf{1} \hat{+} \mathbf{1}$, that originates from the partition 1+1=2, so it must give $\mathbf{2}$:
\be
\mathbf{1} \hat{+} \mathbf{1} = \uparrow \otimes \begin{array}{c}
                 \rc{\uparrow} \\
                 \rc{\downarrow}
               \end{array} =
               \begin{array}{c}
               \uparrow \rc{\uparrow} \\
               \uparrow  \rc{\downarrow}
               \end{array}
               = \mathbf{2} \,.
\ee
Indeed, in the second-last term, the first line is one spin representative of 2, $(\uparrow, \uparrow)$, while the second line is one spin representative of 1+1, $(\uparrow, \downarrow)$. The four-dimensional ($2^N=2^2$), full configuration space, $\mathbf{2}$, is obtained when we spin-flip each final configuration: $(\uparrow, \uparrow)$,$(\downarrow, \downarrow)$ and $(\uparrow, \downarrow)$,$(\downarrow, \uparrow)$. Notice that this is a general feature of this operation: one can consider even just one single representative per each spin-flipped pair of the first term, perform the operation as described earlier, and then, to obtain all the configurations, at the end of the procedure apply the spin-flip.

We are now where we wanted to be. When $\mathbf{N}$ is the configuration space of a Mukhanov-Bekenstein quantum BH, we have found BH self-similarity, in the plain terms we were searching for:

\textit{The configuration space, $\mathbf{N}$, of a BH is made of the configuration spaces of smaller BHs, that in turn are made of configuration spaces of even smaller BHs, and again and again, until we reach $N$ copies of $\mathbf{1}$, the configuration space of the tiniest (elementary) BH.}

Since to any of the $2^{N-1}$ OPIs of $N$ we can associate one of the $2^{N-1}$ OPIs of $\mathbf{N}$
\be
\sum_{i} N_i \ =  \ N \to \hat{\sum}_{i} \mathbf{N_i} = \mathbf{N}  \,, \,\, \sum_{j} M_j = N \to \hat{\sum}_{j} \mathbf{M_j} = \mathbf{N} \,, ... \,,
\ee
where $\hat{\sum}_{i} \mathbf{N_i} = \mathbf{N_1} \hat{+} \mathbf{N_2} \hat{+} \cdots $, whatever pattern we found in the OPI of $N$, it is found in the configuration space $\mathbf{N}$ of the BH, and then repeated for the smaller numbers, until we reach the ``quantum'' of the BH space, $\mathbf{1}$.

A suggestive pattern is given by
\be
\mathbf{N} \ = \ \mathbf{1} \hat{+} (\mathbf{N-1}) \ = \ \mathbf{1} \hat{+} (\mathbf{1} \hat{+}  (\mathbf{N-2})) = \mathbf{1} \hat{+} (\mathbf{1} \hat{+} (\mathbf{1} \hat{+}  (\mathbf{N-3}))) = \cdots
= \hat{\sum}^{N}_{i=1} \mathbf{1} \,.
\ee
Here one can say that when the configuration space of the tiniest BH, $\mathbf{1}$, is isolated from the rest, than this can be repeated again and again till the complete splitting.

It is crucial to notice that, as wanted, in this picture the self-similarity does not require any change of description of the degrees of freedom (e.g., from the evaporating BH to the long string \cite{susskindBOOK}, see also \cite{veneziano1}). What one does there is to find patterns within the configuration space of a given, fixed BH. We are not considering here neither BH \textit{evaporation} nor BHs \textit{merging} \cite{abbotts}.

Let us conclude this part by saying that the constructions of \cite{progress} may solve the problem we started with. On the other hand, they lack any dynamical consideration whatsoever, as only kinematic was the concern there. No configuration is preferred to any other, by the virtue of dynamical properties of the system. In other words, all configurations were treated equally and this can only give back the entropy of (\ref{ClassicalEntropy}), which, with a strong abuse of the language, since we are in a quantum BH model, is sometimes referred to as ``classical entropy''.

This is probably something that will be fully amended only by the long-sought-for final QG theory, see, e.g., \cite{Rovelli:2004tv}, that will tell us how these fundamental (fermionic) degrees of freedom (see, e.g., \cite{aischol,Xons,aismal}) interact, and some, $O(\ln N)$, ``quantum corrections'' have been put forward based on perturbative quantum considerations \cite{Kaul:2000kf, Gupta:2001bg, Ghosh:2004rq, Majhi:2013tw, Singleton:2013ama, Ong:2018xna}.

On the other hand, the simple (simplistic) approach of \cite{progress} has two advantages. First it is based on a non-interacting (free) spin model, that some authors also consider to be a viable candidate \cite{aischol,Xons,aismal}. Second, in order to use an information-theoretical approach, the selection of specific configurations over others is not appropriate. In fact, if a (quantum) BH has indeed to be used as the ultimate (quantum) computer \cite{lloyd}, then one expects all configurations to be treated equally. The actual evolution of the quantum states should not be fixed by a given spin model, but should rather be governed by a specific Hamiltonian that ``implements'' the given ``computation''.

\section{Conclusions}\label{Conclusions}

The interpretation of quark confinement as the effect of a (event) \textit{horizon} for color degrees of freedom, naturally lead to view hadronization as a quantum tunnelling
through such horizon. Taking this view, hadron formation is then the result of a Unruh phenomenon, related with the string breaking/string formation mechanism. This is so because the large-distance QCD potential generates a constant and enormous \textit{acceleration}, $a \simeq 3.2 \times 10^{33} {\rm m}/{\rm s}^2$, that is precisely what we need to have a measurable Unruh effect, $T_U \sim 4 \times 10^{11} {\rm K} \sim 170 {\rm MeV}$.

This opens the way to a clear explanation of the thermal behaviors on both arenas, hadron physics and BH physics. For instance, this immediately explains why the hadronization temperature, $T_h$, is universal when seen as a $T_U$. Indeed, $T_h$ is found to be the same for small and large initial collision settings, whereas $T_U$ is fixed, once and for all, by the value of the acceleration, $a$. This also explains why hadrons are born in equilibrium.

In fact, the Hawking-Unruh radiation is an example of a \textit{stochastic} rather than \textit{kinetic} equilibrium. The reason behind the randomization are not repeated and casual collisions among particles, but rather the quantum entanglement between degrees of freedom on the two sides of the barrier to information transfer, that is the event horizon. The temperature is then determined by the strength of the ``confining'' field.

In the chromodynamical counterpart of this phenomenon, described in this review, the ensemble of all produced hadrons, averaged over all events, leads to the same equilibrium distribution as obtained in hadronic matter by kinetic equilibration. For a very high energy collision, with a high average multiplicity, even one event alone can provide such equilibrium. The destruction of memory, which in kinetic equilibration is achieved through sufficiently many successive collisions, is here automatically provided by the tunnelling process.

The above are the physical fundamental aspects common to both types of phenomena. On this the analogy can be solidly established, and many results can be obtained: The string breaking and the BH entropy analogy, that reproduces the ``magic numbers'' characterizing the freeze-out curve; The strangeness production, at low parton density, that is due to different Unruh temperatures in the single string breaking; At high energy and  multiplicity, the large parton density, in the transverse plane, that removes the different temperatures, by string (or color event horizon) overlap, giving the strangeness enhancement; The self-similar behaviour, characteristic of the hadronic production, then drives research on the self-similarity of BH configurations.

Let us then close on an optimistic note, by saying that this somehow new and original analog system of QG has many other results to grasp. In particular, given the richness of the thermodynamical aspects of the SHM, it could lend itself to probe the dynamical aspects of BH evaporation that are the next frontier of this field.

\section*{Acknowledgements}

P.C. and A.I. were supported by Charles University Research Center (UNCE/SCI/013). P.C. gladly acknowledges the kind hospitality of the Institute of Particle and Nuclear Physics of Charles University. A.~I. thanks the INFN, Sezione di Catania, for supporting a visit to Catania University, where this work was completed.


\bibliographystyle{apsrev4-2}
\bibliography{Bibliografia_Universe}

\end{document}